\documentclass[prb,twocolumn,preprintnumbers,amsmath,amssymb]{revtex4}
\usepackage{graphicx}
\usepackage{color}
\usepackage{multirow}
\usepackage[sort&compress]{natbib}

\newcommand{\ket}[1]{\ensuremath{| #1 \rangle}}
\newcommand{\meV}{\ensuremath{\,\mbox{meV}}}

\newcommand{\nm}{\ensuremath{\,\mbox{nm}}}

\newcommand{\vc}[1]{\ensuremath{\vec{#1}}}

\newcommand{\Hop}{\ensuremath{\mathcal{H}}}

\newcommand{\Pth}{\ensuremath{P_{th}}}

\begin{document}

\title{Configuration interaction calculations of the controlled phase gate in double quantum dot qubits}
\author{Erik Nielsen, Richard P.~Muller and Malcolm S.~Carroll}
\affiliation{Sandia National Laboratories, Albuquerque, New Mexico 87185 USA}
\date{\today}

\begin{abstract}
We consider qubit coupling resulting from the capacitive coupling between two double quantum dot (DQD) single-triplet qubits.  Calculations of the coupling when the two DQDs are detuned symmetrically or asymmetrically are performed using a full configuration interaction (CI).  The full CI reveals behavior that is not observed by more commonly used approximations such as Heitler London or Hund Mulliken, particularly related to the operation of both DQDs in the (0,2) charge sector. We find that there are multiple points in detuning-space where a two-qubit entangling gate can be realized, and that trade-offs between coupling magnitude and sensitivity to fluctuations in detuning make a case for operating the gate in the (0,2) regime not commonly considered.
\end{abstract}

\maketitle

The lowest energy singlet (spin zero) and triplet (spin one) states of two spins in a double quantum dot (DQD) form a common qubit encoding choice for semiconductor based quantum computing.\cite{PettaScience_2005} By choosing these states as the qubit computational basis, quantum information is stored in the spin of the wave function.  This has the advantage of being less susceptible to charge noise when compared to charge qubits,\cite{Tanamoto_2000} where the two computational basis states have very different electronic charge distributions (\emph{e.g.}, when a single electron in the left and right dot constitute the computational basis).

Two qubit coupling is a necessary element for implementing quantum computing.  Approaches for coupling two (or more) DQD qubits have almost universally utilized the Coulomb interaction between neighboring DQDs.\cite{TaylorNatPhys_2005}  We refer to the resulting gate, which is equivalent to a controlled phase (CPHASE) gate up to single qubit rotations, as the ``Coulomb gate''.  The Coulomb gate is constructed by simply coupling two DQDs capacitively, but not allowing electrons to tunnel between them.\cite{StepanenkoBurkard_2007}  This presents a significant challenge because, for the Coulomb gate to operate the singlet and triplet states of each qubit must have different electronic charge distributions, effectively turning each DQD into a \emph{charge qubit} and making the system more vulnerable to charge noise.  This paper examines the susceptibility of the Coulomb gate to voltage fluctuations using a more quantitative and qualitative method, full CI, than previous approaches.  We also highlight a more robust gating sequence to minimize the susceptibility.





\begin{figure}
\includegraphics[width=2.2in,angle=270]{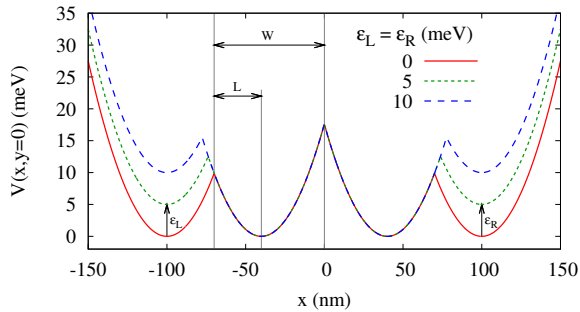}
\caption{Potential energy slice at $y=0$ of the four-dot system given by Eq.~\ref{eqPot}.  The dot half-separation $L$ (of a single DQD), and the half-separation $W$ of the DQDs are labeled, as well as the detuning energies $\epsilon_L$ and $\epsilon_R$ of the left and right DQDs, respectively. Here $L=30\nm$ and $W=70\nm$. Curves for symmetrically detuned cases $\epsilon_L = \epsilon_R = 0$, $5$, and $10\meV$.\label{figFourDotPotential}}
\end{figure}

\textbf{Model.}
We model a double DQD system potential by the minimum of four parabolic dots.  The curvature of each dot is identical, and all four dots lie along the $x$-axis.  We consider two-dimensional systems, a good approximation for lateral dot structures where either a heterojunction or electric field strongly confines the electrons in the vertical direction.  The single particle Hamiltonian, in the absence of magnetic field, is given by
\begin{equation}
\Hop^{1P} = \frac{p^2}{2m^*} + V(\vc{r})\label{eq1PHam}
\end{equation}
where
\begin{equation}
\begin{array}{ll}
V(x,y) = & \frac{1}{2}m^*\omega^2 \left[ y^2 + \min\left( \right.\right. \\
& (x-L-W)^2+\epsilon_L,\, (x+L-W)^2, \\
& \left.\left. (x-L+W)^2,\, (x+L+W)^2+\epsilon_R \right)\right]\,. 
\end{array}
\label{eqPot}
\end{equation}
The first and second parabolic potentials being minimized (second line of Eq.~\ref{eqPot}) form the first DQD, while the third and fourth (third line) form the second DQD.  The two dots of a single DQD are separated by length $2L$ and the centers of the two DQDs are separated by length $2W$ (see Fig.~\ref{figFourDotPotential}).  Detuning is modeled as raising the potential of the \emph{outer} dot: the left dot of the first (left) DQD by $\epsilon_L$ and the right dot of the second (right) DQD by $\epsilon_R$.  The full many-electron Hamiltonian

\begin{equation}
\Hop = \sum_{i=1}^N \Hop_i^{1P} + \sum_{i<j} \frac{e^2}{\kappa r_{ij}} 
\label{eqMBHam}
\end{equation}
for $N$ electrons, where $\Hop_i^{1P}$ is the single particle Hamiltonian of Eq.~\ref{eq1PHam} for the $i^{th}$ particle, $\kappa$ is the material dielectric constant, and $r_{ij} = |\vc{r}_i-\vc{r}_j|$.

Hamiltonian $\Hop$ is solved by a full configuration interaction (CI) technique detailed in Ref.~\onlinecite{Nielsen_ImplicationsExchange_2010}.  This CI implementation uses $s$-type Gaussian function to form the single-particle basis from which we obtain the multi-electron eigenenergies and eigenstates of the system, as well as their spin eigenvalues. Due to the larger number of electrons and dots compared to a single DQD (4 instead of 2 in both cases), computational restrictions prevent the double-DQD system from being run with as many Gaussian basis elements per dot, and thus the convergence of the method is worse than for that of a single DQD.  In the results that follow, we use five Gaussian functions per dot, arranged in a plus-sign (+) pattern and do not variationally improve the basis.  Consequently, we treat the results as only qualitatively accurate.  We do not consider the basis size limitation a significant shortcoming at this level, however, since in any realistic device the environmental couplings and device geometry will affect the quantitative results much more than the approximation assumed by using a limited basis.  A primary strength of this approach is that is reveals behavior that is not observed in simpler approximations such as Heitler-London and Hund-Mulliken, which are commonly used for DQD qubit analysis.  In all the calculations below, we use GaAs material parameters $m^* = 0.067\,m_e$, $\kappa = 12.9$, and $g=2$, and the DQD parameters $E_0=5\meV$ and $L=30\nm$.

\textbf{Single DQD.}
First, we review the charge configuration dependence of a DQD on detuning, which assists the description of the Coulomb gate in following sections.  When the inter-dot barrier is large (compared to the confinement energy) the electronic state can understood as a superposition of basis states, each of which lies in a particular charge sector.  Charge sectors are labeled by pairs of integers $(n,m)$ which indicate that there are $n$ electrons in the left dot and $m$ in the right dot.  Thus, a two-electron DQD wave function can be in a superposition of the (1,1), (2,0), and (0,2) charge sectors.

\begin{figure}
\includegraphics[width=2.2in,angle=270]{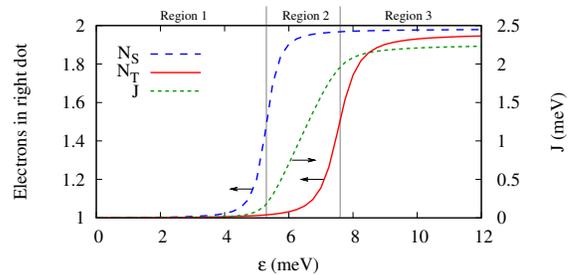}
\caption{The electron occupation of the right dot and the exchange energy $J$ as a function of detuning $\epsilon$ for a single DQD. Vertical lines mark the points where the singlet and triplet states have exactly 1.5 electrons in the right dot (the middle of the transition between the (1,1) and (0,2) charge sectors), and divide the domain into three regions as labeled at the top of the plot.  Parameters $E_0=5\meV$, $L=30\nm$. \label{figDQD}}
\end{figure}

We consider a DQD with identical parabolic dots, so that the Hamiltonian is given by Eq.~\ref{eqMBHam} except the potential found in the single particle Hamiltonian is 
\begin{equation}
V(x,y) = \frac{1}{2}m^*\omega^2 \left[ \min\left( (x-L)^2+\epsilon, (x+L)^2 \right) + y^2\right] \label{eqSingleDQDPot}
\end{equation}
instead of Eq.~\ref{eqPot}.  There are only two parameters, $L$ and $\epsilon$ which determine the shape of this potential by setting (half) the distance between the dots and the detuning, respectively.  When the inter-dot separation is fixed and $\epsilon$ varies from zero to some positive energy, the lowest lying two-electron singlet and triplet states both transition from the (1,1) to the (0,2) charge sector, but do so at different values of the detuning, as shown in Fig.~\ref{figDQD}.  The singlet state typically transitions at lower detuning because in the (0,2) singlet wave function the spin antisymmetry allows both electrons to occupy the lowest energy combination of single-electron orbitals, while this is not true of the triplet.  Transitions of the singlet and triplet divide the range of $\epsilon$ into three regions, labeled along the top of Fig.~\ref{figDQD}. In regions 1 (3), both singlet and triplet states are in (1,1) ((0,2)) charge sector. In region 2 the singlet is a (0,2)-state while the triplet is a (1,1)-state, essentially making the system into a charge qubit and giving the strongest potential for capacitively coupling the qubit to another qubit (or to a charge measurement device). Figure \ref{figDQD} also shows the exchange energy $J$ defined as the energy difference between the triplet and singlet states and proportional to the speed of the qubit's rotation.  A study of the exchange energy in DQDs using the CI method can be found in Ref.~\onlinecite{Nielsen_ImplicationsExchange_2010}.

\textbf{Double DQD}
Two double quantum dots can be coupled capacitively so that differences in the charge distribution within one DQD affect the electrons in the other DQD, and vice versa.  When each DQD contains two electrons and is used as a singlet-triplet qubit, the capacitive coupling gives rise to the two-qubit Coulomb gate.  Let $\ket{S}$ and $\ket{T}$ denote the lowest lying singlet and triplet states, respectively, of a single DQD.  The effective Hamiltonian of an ideal DQD, in the absence of magnetic field gradients, in the basis $\left\{ \ket{S},\ket{T} \right\}$ is
\begin{equation}
\Hop = \left[
\begin{array}{cc}
0 & 0 \\
0 & J
\end{array}
\right]
\end{equation}
where $J$ is the exchange energy (plotted in Fig.~\ref{figDQD} as a function of the detuning $\epsilon$).  Similarly, the effective Hamiltonian of two capacitively coupled DQDs (without a magnetic field gradient) in the basis $\left\{ \ket{SS},\ket{ST},\ket{TS},\ket{TT} \right\}$ will in general be
a diagonal matrix with elements $\{ E_{SS}, E_{ST}, E_{TS}, E_{TT} \}$,
where we introduce the labels $E_{SS}$, etc., for the energies of the respective states.  The matrix is approximated as diagonal because the Coulomb interaction is assumed not to couple different spin subspaces, and we assume that electrons do not tunnel between the DQDs.  The degree of control one qubit exerts on the other is given by the quantity
\begin{equation}
\Delta = E_{TT} - E_{TS} - E_{ST} + E_{SS}
\end{equation}
which can be understood as the difference of the left DQD's exchange energy when the right DQD is in the singlet and the triplet state, or vice versa.  That is, $\Delta = J_T - J_S$ where $J_S \equiv E_{TS} - E_{SS}$ and $J_T \equiv E_{TT} - E_{ST}$.  Thus, $\Delta$ gives a measure of how much control the state of one qubit has on the rotation rate of the other.  Similar to the exchange energy $J$ corresponds to the speed of single qubit rotation, the value of $\Delta$ is the difference in one (say the left) qubit's rotation speed when the  other (right) qubit moves between $\ket{0}$ and $\ket{1}$, and thus corresponds to speed of two-qubit \emph{controlled} rotation.

\begin{figure}
\includegraphics[width=2.2in,angle=270]{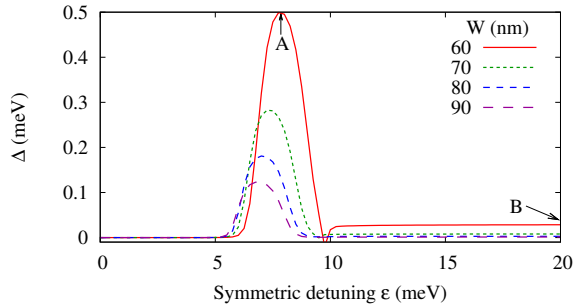}
\caption{Two-qubit coupling parameter $\Delta$ as a function of the symmetric detuning $\epsilon$ of two DQDs.  Parameters $E_0=5\meV$ and $L=30\nm$.  Curves for different $W$ are shown to illustrate the effect of different capacitive couplings, and possible operating points A and B (see text) are labeled on the $W=60\nm$ curve.\label{figSymCoulombGate}}
\end{figure}

\emph{Symmetric detuning.}
While the two DQDs comprising the gate can be detuned independently, first consider the ``symmetric detuning'' case where $\epsilon_L = \epsilon_R \equiv \epsilon$, so that the potential of both outer dots are raised by the same amount $\epsilon$.  As $\epsilon$ is increased, each DQD will move through the three regions of Fig.~\ref{figDQD}.  Figure \ref{figSymCoulombGate} shows how $\Delta$ varies with $\epsilon$ for a typical double-DQD system.  The region of large $\Delta$ corresponds to region 2 of Fig.~\ref{figDQD} where each DQD behaves most like a charge qubit, and is the regime most often considered for the Coulomb gate's operation.  We refer to this range of $\epsilon$ as the ``large-$\Delta$'' region of the double-DQD system.  Figure \ref{figSymCoulombGate} also shows that by increasing the inter-DQD distance $W$ (decreasing the capacitance), the magnitude of $\Delta$ lessens, as expected.  These distances correspond to capacitances of several atto-Farads. 

The primary contention of this paper is that the large-$\Delta$ region is not the only region in which the Coulomb gate may be operated, and that there are important trade-offs to consider when choosing where to operate the Coulomb gate.  One metric is the strength of the two-qubit coupling $|\Delta|$, which one would like to be large so there is sufficient signal and so the gate is fast relative to charge dephasing mechanisms in the environment.  A second metric is sensitivity to voltage fluctuations such as electronics (or ``control'') noise.  The sensitivity depends on the average slope of the $\Delta$ vs.~$\epsilon$ curve over some uncertainty window $d\epsilon$.  Define points A and B on a $\Delta$ vs. $\epsilon$ curve to be the points at maximal $\Delta$ and at large $\epsilon$, respectively.  For the curves of Fig.~\ref{figSymCoulombGate} we set point B at $\epsilon=20\meV$, and find for each curve that, although operating at point B gives smaller $|\Delta|$, it also is much less sensitive to fluctuations in $\epsilon$.  Quantitatively, Table \ref{tableTradeoffs} compares points A and B for each curve in Fig.~\ref{figSymCoulombGate}, the $\pi$-controlled-rotation gate (similar to CNOT) time $t_\pi = \pi\hbar/|\Delta|$ and the maximum allowable error in $\epsilon$ which achieves a gate error probability less than a given threshold value $\Pth$, which we set as $10^{-4}$ in our comparisons below.  For the parameters we have used, sub-nanosecond gate times result when operating at the peak, which are beyond current electronics capability.\cite{LevyClassicalConstraints_2009} Thus, in this model, one must reduce the capacitance between the dots to operate at the peak in $\Delta$ with nanosecond or longer gate times.  The maximum allowable detuning error, $d\epsilon$, is $2-6$ times larger when operating the gate on the flat region of the $\Delta$ curve at $\epsilon=20\meV$.  This indicates that that the flatness of the curves in Fig.~\ref{figSymCoulombGate} at large $\epsilon$  more than compensate, from a gate error perspective, for the decrease in gate speed.



\begin{table}
\begin{tabular}{|c|c|c|c|c|}
\hline
W & \multicolumn{2}{|c|}{$t_\pi$ (ns)} & \multicolumn{2}{|c|}{$d\epsilon$ (meV)} \\ 
(nm) & peak & $\epsilon=20\meV$ & peak & $\epsilon=20\meV$ \\ \hline \hline 
60 & 0.03 & 1   & 0.15  & 0.91  \\  \hline
70 & 0.05 & 5   & 0.16  & 0.73  \\  \hline
80 & 0.07 & 40  & 0.12  & 0.40  \\  \hline
90 & 0.1  & 500 & 0.16  & 0.32  \\  \hline
\end{tabular}
\caption{ Gate time ($t_\pi$) and maximum allowable detuning error ($d\epsilon$) required to achieve the gate error threshold $\Pth=10^{-4}$.  Values are given for operating the gate at the peak and $\epsilon=20\meV$ points (A and B in text) of each curve in Fig.~\ref{figSymCoulombGate}. \label{tableTradeoffs}}
\end{table}

The trade-off between the magnitude of $\Delta$ and lower gate error probability suggests that operating in the large-$\epsilon$ region (point B) would be advantageous when a primary source of noise is fluctuations in the biasing voltages (\emph{e.g.}, overshoot and/or ringing).  



\begin{figure}
\includegraphics[width=2.1in,angle=270]{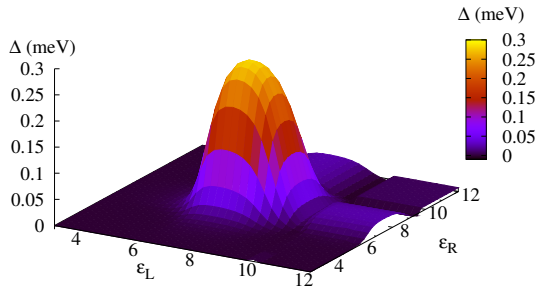}
\caption{Qubit coupling metric $\Delta$ as a function of $\epsilon_L$ and $\epsilon_R$.  The lower pane shows a projected 2D color plot of the same data as the upper plot.  Parameters $E_0=5\meV$, $L=30\nm$, and $W=70\nm$ were chosen so the features are more easily visible.\label{figAsymCoulombGate}}
\end{figure}

\emph{Asymmetric detuning.}
We now consider the case when $\epsilon_L$ and $\epsilon_R$ are varied independently.  Such capability will already be a requirement for most implementations, since in order to perform single qubit gates the detuning of each DQD must be accessible separately. Figure \ref{figAsymCoulombGate} shows the behavior of the two-qubit coupling strength $\Delta$ as a function of $\epsilon_L$ and $\epsilon_R$ as a surface in three dimensions and also projected in 2D as a color plot.  $\Delta$ is significantly larger within the square $\{ (\epsilon_L,\epsilon_R) \mathrm{s.t.} \epsilon_L \in [\epsilon^*_S,\epsilon^*_T] \mathrm{ and } \epsilon_L \in [\epsilon^*_S,\epsilon^*_T] \}$, where $\epsilon^*_S$ and $\epsilon^*_T$ are the detuning values at which the singlet and triplet, respectively, transition from the (1,1) to the (0,2) charge sector.  With the parameters of Fig.~\ref{figAsymCoulombGate}, $\epsilon^*_S \approx 6\meV$ and $\epsilon^*_T \approx 9\meV$.

The shape of the surface in Fig.~\ref{figAsymCoulombGate}, which is qualitatively unchanged for different dot separations and sizes, illustrates several points.  First, there are several different regimes in which the double-DQD system can be biased to achieve a nontrivial two-qubit gate - all that is required is that $|\Delta| > 0$.  We find essentially the same two potential operating points as in the symmetric-detuning case: the peak at $\epsilon_L=\epsilon_R\approx 7\meV$ and the plateau at large $\epsilon$.  Figure \ref{figAsymCoulombGate} illustrates the typical behavior that the peak is equally or more sharp along the $d\epsilon_L=-d\epsilon_R$ direction as it is along the symmetric $d\epsilon_L=d\epsilon_R$ direction.  Operating at the peak is more sensitive to both types of control noise than operating on the plateau.

\begin{figure}
\begin{tabular}{cc}
\hspace{-0.3in}\raisebox{-0.8cm}{\includegraphics[width=1.6in,angle=270]{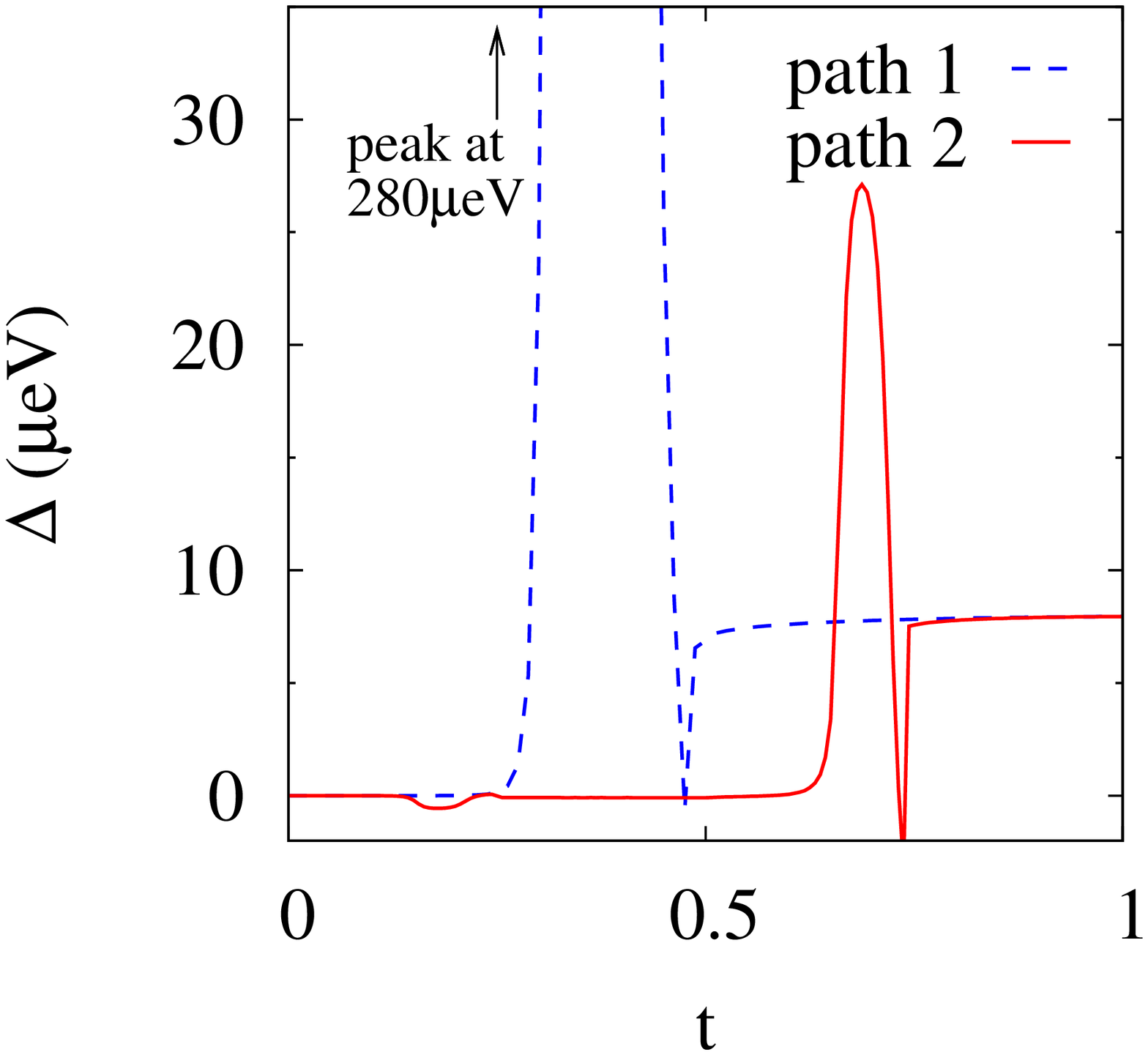}} & 
\hspace{-0.7in}\includegraphics[width=2.0in,angle=270]{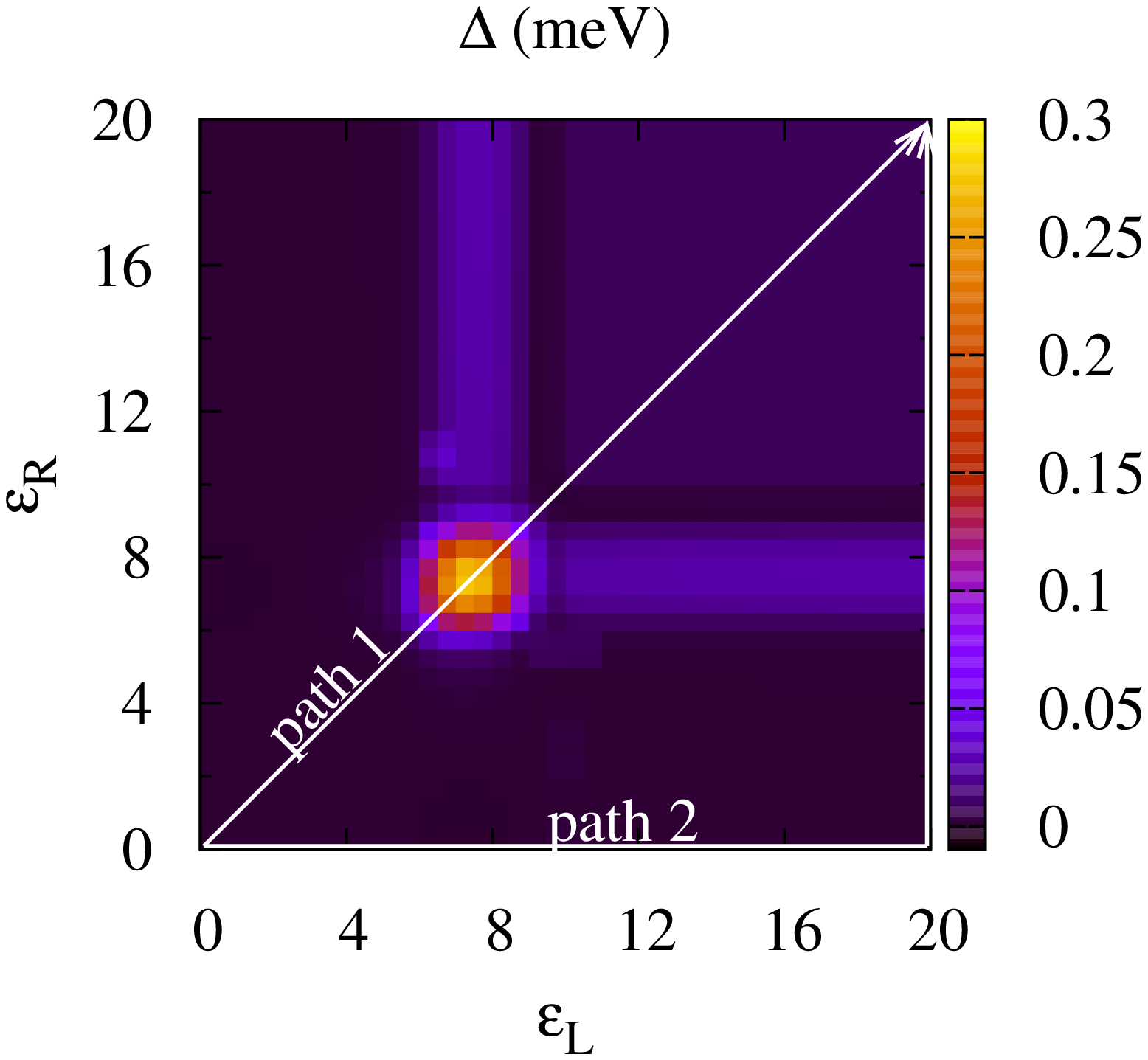} \\
\end{tabular}
\caption{ $\Delta$ as the detuning $(\epsilon_L,\epsilon_R)$ is varied along two paths.  Along path 1, $\epsilon_L = \epsilon_R = 20t\meV$.  Along path 2, $\epsilon_L$ is first changed from $0$ to $20\meV$ while $\epsilon_R=0\meV$, then $\epsilon_R$ is changed from $0$ to $20\meV$ while $\epsilon_L=20\meV$.   \label{figPathContrast}}
\end{figure}

If one does operate the gate on the large-$\epsilon$ plateau, any path in $\epsilon_L$-$\epsilon_R$ ``detuning space'' may be chosen to move between the idle and controlled rotation detuning points.  If the idle point is $\epsilon_L = \epsilon_R = 0$ and the rotation point at $\epsilon_L = \epsilon_R = \mbox{large}$ (say $20\meV$ for the example in Fig.~\ref{figAsymCoulombGate}), then it is advantageous to follow the path which detunes the left, then the right, DQD instead of detuning both DQDs symmetrically (cf. Fig.~\ref{figPathContrast}).  This avoids the region of large $\Delta$ which is sensitive to control errors, and thus will result in a gate which is more robust to control noise and easier to characterize.  


\textbf{Conclusion.}
We have investigated the operation of a two-qubit gate resulting from capacitively coupled singlet-triplet GaAs double quantum dot qubits.  The strength of the two-qubit coupling is obtained both for symmetric detuning of the two DQDs and for the case where the detuning of the left and right DQDs is different.  We find that, similar to our work on the single-qubit exchange gate,\cite{Nielsen_ImplicationsExchange_2010} there exists a regime where the gate operation is relatively insensitive to fluctuations in the DQD detuning.  Furthermore, we show biasing the DQDs non-symmetrically may allow this regime to be reached without moving through a region where the gate is sensitive to bias fluctuations.  This work suggests that for DQDs with sufficient capacitive coupling, operating the Coulomb gate in a large-$\epsilon$ regime will mitigate adverse effects of control voltage fluctuations.  The full configuration interaction method used shows features not visible to Heitler-London and Hund Mulliken techniques.

This work was supported by the Laboratory Directed Research and Development program at Sandia National Laboratories. Sandia National Laboratories is a multi-program laboratory managed and operated by Sandia Corporation, a wholly owned subsidiary of Lockheed Martin Corporation, for the U.S. Department of Energy's National Nuclear Security Administration under contract DE-AC04-94AL85000.


\bibliography{cphasePaper}

\end{document}